\documentclass[11pt,twoside]{article}  % Leave intact
\usepackage{adassconf}
\begin{document}   % Leave intact
\paperID{O4-1a}
\title{Hyperatlas: A New Framework for Image Federation}
\author{R.\ D.\ Williams, S.\ G.\ Djorgovski, M.\ T.\ Feldmann, and J.\ C.\ Jacob}
\affil{California Institute of Technology, Pasadena, California}
\contact{Roy Williams}
\email{roy@cacr.caltech.edu}
\paindex{Williams, R.D.}
\aindex{Djorgovski, S.G.}
\aindex{Feldmann, M.T.}
\aindex{Jacob, J.C.}
\authormark{Williams, Djorgovski, Feldmann, \& Jacob}

\begin{abstract}          % Leave intact
Hyperatlas is an open standard intended to facilitate the large-scale federation of image-based data. The subject of hyperatlas is the space of sphere-to-plane projection mappings (the FITS-WCS information), and the standard consists of coherent collections of these on which data can be resampled and thereby federated with other image data. We hope for a distributed effort that will produce a multi-faceted image atlas of the sky, made by federating many different surveys at different wavelengths and different times. We expect that hyperatlas-compliant imagery will be published and discovered through an International Virtual Observatory Alliance (IVOA) registry, and that grid-based services will emerge for the required resampling and mosaicking.
\end{abstract}
\section{Introduction}
In this paper we discuss a new and largely unexplored data type in
astronomy: large-scale federated imagery. By this, we mean a set of 
sky images that have been resampled (reprojected) to the same pixel
space. The set of images might represent the sky at different
wavelengths, times, resolution, polarization, etc. Because they are on the same pixel plane, these resampled images can be stacked and mined together -- they can be {\it federated}.
The benefits of such  
federation can outshine a possible loss of data quality from resampling.
We will describe scientific possibilities of image federation, define the meaning of an {\it atlas}, then describe a proposed set of standard atlases (Hyperatlas) that will enable interoperability of images that have been resampled by different groups.

When a sky survey is made, the telescope images different parts of the
sky at different times, wavebands, etc, creating a mosaic of
images. Traditionally, the images are converted to star/galaxy
catalogs through source-extraction, and much of the scientific
knowledge is mined from these catalogs. Catalogs are federated by cross-matching, and further knowledge mined from the joint catalog.
However, the intrinsic limitation of any source extraction process is that object
detection relies on some filter, which is usually optimized for a particular source size and/or
contrast, and may not be efficient in detecting useful signals which fall outside its 
parameter range; an example are low surface brightness objects.  Thus, images may contain
valuable information not present in the source catalogs.

In this paper, we consider a new paradigm for mining knowledge from
the images of the sky surveys: by federating the images directly to a
uniform pixel space, so that multiple image sets can be stacked, then doing source detection, pattern matching and other data mining in the resulting multi-channel image. If the pixel spaces ({\it pages}) are standardized, then it is much easier to federate data made by different groups in different ways.

\section{Federation}

Even with an "optimal" source extraction, in some cases even the absence of a
statistically significant signal represents a useful information, but $only$
in the context of other imagery; an example are high-redshift objects detected
with the "dropout" techniques.
Image federation is thus useful in the detection of faint
sources [Szalay 1998], which may not be detected in a particular (or any given)
channel, but may be detected in others; or, a collective of individually marginal
detections may have a high significance when considered jointly.  
We can go fainter in image space because we have more photons from the combined
images and because the multiple
detections can be used to enhance the reliability of sources at a given threshold.

Astronomers are more and more reaching into the time domain
(synoptic), imaging the sky repeatedly for discovery of transient and
variable sources. Near-Earth asteroids and Kuiper-belt objects,
cataclysmic variables, active galactic nuclei, and quasars are all
interesting, as well as lensing events, orphan afterglows, and most
exciting would be a rich set of new types of transients. Image
federation shows promise for analysis of this kind of data, by forming
an average sky, a maximal brightness sky, a sky specialized to
non-Gaussian noise sources, and a sky where information from dozens of
channels is represented in a single color image. Very faint periodic sources could be found by
time autocorrelation of pixel brightness.

A prerequisite for finding differences between sky images is the
perfect mutual positioning of sources.  We
will focus on reconciling images with very different inherent spatial
resolution by micro-optimization of astrometry through a fitting
process. More generally, we can subtract out ``well-fitted'' objects
(e.g., PSF-like), to leave only the faint, complex structure that may
contain new knowledge.

Image federation enables robust detection and flux measurement of
complex, extended sources over a range of size, wavelength, and time
scales. Larger objects in the sky may have both
extended structure (requiring image mosaicking) and a much smaller
active center, or diffuse structure entirely. We will be able to
combine multiple instrument imagery to build a multi-scale,
multi-wavelength picture of such extended objects. We will also
make statistical studies of
extended, complex sources that vary in shape with wavelength.

\section{Pages and Atlases}
Federated imagery is a collection of mosaicked image products, all on the same pixel plane -- we call these {\it channels}. More precisely:

\begin{itemize}
\item We define a {\bf Page} to be a smooth, continuous, invertable
map between the celestial sphere and a plane, with a ``pointing center'', 
at which Jacobian of the map is unit, meaning free of distortion. A
Page is just the specification of the projection: it is semantically
equivalent to the WCS keywords CRVAL (pointing center), CTYPE
(projection type), CDELT (scale), and CD/CROTA (rotation matrix). A page may have a pixel plane of infinite extent, for example the TAN projection stretches to infinity.

\item We define a {\bf Plate} to be a rectangular area on the plane of
a Page together with single-channel pixel data filling that rectangle,
and each pixel of the plate having the same data types as the FITS
standard. Thus a Plate is semantically equivalent to a single-plane
FITS image. Besides the data content, a Plate adds the WCS keywords
NAXIS (numbers of pixels), CRPIX (shift from the pointing center).

\item An {\bf Atlas} is a coherent collection of Pages. Generally all
pages in an atlas have the same scale and projection, and the pointing
centers are laid out rationally, perhaps to cover the whole sky,
perhaps just a particular region of interest like the galactic plane.

\item We define a {\bf Channel} to be a uniform collection of Plates that is rendered to a given atlas, so that there is at most one pixel for any position in the sky -- ie. a monochrome covering. In general there will be overlap of plates -- the same point of the sky shown on multiple pages -- in which case the channel-values of that point mus be the same independent of which page it is on. The channel could represent fluxes in different wavelength bands, times, polarizations, etc, or it may have been derived from other channels by computation.
\end{itemize}

\section{Hyperatlas Standard}
In the interests of interoperability, we have specified a
number of standard atlases and called the collection the {\it Hyperatlas} standard. For more information see [Williams 2003a].
We hope that large surveys and vital data can all
be resampled to the same set of pages, meaning that their images are
on exactly the same pixel grid, and can be compared
directly.

Part of the atlas definition is the collection of pointing centers of
its pages. In Hyperatlas, there are currently two choices for this
layout: TM and HV. The former puts pointing centers on lines of
constant declination, the latter on the centers of Hierarchical
Triangular Mesh (HTM) [Kunzst 1999] triangles. Each layout comes with a
parameter: for the TM layout, the parameter means the separation
between the declination lines; and for the HV layout it is the level of the HTM hierarchy. In the figure is an atlas made from the
TM-5 configuration of pointing centers. 
Each of these layout types is defined by an algorithm or pseudocode (see [Williams 2003]), and also available through redundantly-implemented web services. 

The atlas is further specified by the projection type (SIN, TAN, AIT,
etc.\ from the WCS library), and by the nominal scale at
the pointing center. We have discretized the scale numbers in powers
of two: scale 20 is one arcsecond per pixel, and generally scale $S$
is $2^{20-S}$ arcseconds per pixel.

For example the atlas {\bf TM-5-TAN-20} consists of 1734 pages
covering the celestial sphere, with every point of the sphere at
most 3.5 degrees (= $5/\sqrt 2$) from a pointing center, a scale of 1
arcsecond per pixel, and the projection TAN.
\begin{figure}
\epsscale{.40}
\plotone{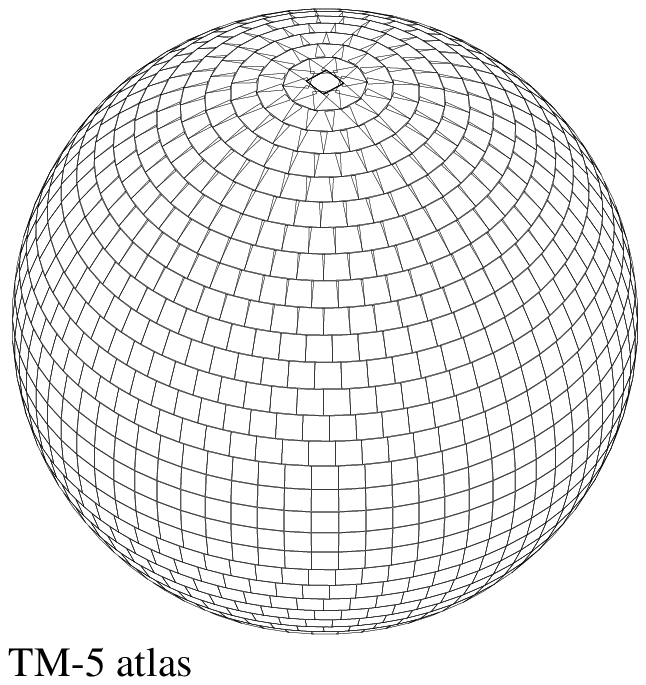}
\caption{\small A layout of pages where every point of the celestial sphere is at most 3.5 degrees from a pointing center. In this illustration, each page is illustrated with a plate 5 degrees on a side.} \label{arch}
\end{figure}
Hyperatlas is not only documents, but is also implemented as web services [Williams 2003]. There is a service to provide all the pages of an atlas, or specific page metadata by number or by sky position.

Recently, the International Virtual Observatory Alliance (IVOA) agreed on
a flexible and extensible framework for registering astronomical resources in a distributed virtual registry [IVOA 2003]. We expect to enhance the Hyperatlas standard by providing an extension schema for this registry. This would allow Hyperatlas-compliant imagery to be published and discovered through the Virtual Observatory, bringing to full flower the promise of image federation technology.
\acknowledgments
We are grateful to the National Science Foundation for support of this
work through the National Partnerships for Advanced Computational
Infrastructure and through the National Virtual Observatory project. SGD thanks the NASA AISRP program for support. JCJ thanks NASA for his support.

% Do not place any material after the references section


\begin{references}
\reference IVOA, International Virtual Observatory Registry Framework,  2003, \\
{\tt \small http://www.ivoa.net/twiki/bin/view/IVOA/IvoaResReg}
\reference Szalay, A.\ S., Connolly, A.\ J., Szokoly, G.\ P., Simultaneous Multicolor  Detection of Faint Galaxies in the Hubble Deep Field,  \aj, 117, 68, also  astro-ph/9811086.
\reference Williams, R.\ D., The NVO Hyperatlas Standard, 2003, \\{\tt \small
http://bill.cacr.caltech.edu/usvo-pubs/files/hyperatlas.pdf}
\reference Williams, R.\ D., Feldmann, M.\ T., Atlasmaker, Software to build Atlases of  Federated Imagery, 2003, \\{\tt \small
http://www.cacr.caltech.edu/projects/nvo/atlasmaker/}
\end{references}
\end{document}